\documentclass[aps, prl, twocolumn, superscriptaddress, showpacs]{revtex4}
\usepackage{amssymb}
\usepackage{amsmath}
\usepackage{bm}
\usepackage[dvips]{graphicx}
\usepackage{dcolumn}
\usepackage{longtable}

\begin{document}

\title{Anisotropic magnetic fluctuations in the ferromagnetic superconductor UCoGe studied by angle-resolved $^{59}$Co NMR}
\author{Y.~Ihara}
\altaffiliation{ihara@scphys.kyoto-u.ac.jp}
\author{T.~Hattori}
\author{K.~Ishida}
\author{Y.~Nakai}
\affiliation{Department of Physics, Graduate School of Science, Kyoto University, Kyoto 606-8502, Japan}
\author{E.~Osaki}
\author{K.~Deguchi}
\author{N.~K.~Sato}
\affiliation{Department of Physics, Graduate School of Science, Nagoya University, Nagoya 464-8602, Japan}
\author{I.~Satoh}
\affiliation{Institute for Materials Research, Tohoku University. Sendai 980-8577, Japan}
\date{\today}

\begin{abstract}
We have carried out direction-dependent $^{59}$Co NMR experiments on a single crystal sample of the 
ferromagnetic superconductor UCoGe in order to study the magnetic properties in the normal state.  
The Knight shift and nuclear spin-lattice relaxation rate measurements provide microscopic evidence that 
both static and dynamic susceptibilities are ferromagnetic with strong Ising anisotropy. 
We discuss that superconductivity induced by these magnetic fluctuations prefers spin-triplet pairing state. 
\end{abstract}

\pacs{74}
\maketitle

Superconductivity near ferromagnetism, found in several U-based compounds
\cite{saxena-Nature406,aoki-Nature413,akazawa-JPCM16,huy-PRL99}, 
often exhibits intriguing superconducting (SC) properties under magnetic fields, 
such as the extremely high upper critical field ($H_{c2}$) \cite{slooten-PRL103,aoki-JPSJ78} 
and the reentrant superconductivity in large external magnetic fields 
\cite{levy-Science309, miyake-JPSJ77}.  
SC pairing mechanisms of ferromagnetic (FM) superconductivity cannot be explained 
in the framework of  the conventional Bardeen-Cooper-Schrieffer theory. 
Many experimental and theoretical studies suggest that 
magnetic fluctuations near critical points give rise to attractive electron-electron interactions 
to form Cooper pairs in unconventional superconductors. 
The normal-state magnetic properties must be investigated 
in order to understand the diverse properties of FM superconductivity. 

Theoretically studies, taking itinerant ferromagnets, e.g. ZrZn$_{2}$ and Ni metal, 
as model materials \cite{fay-PRB22}, predicted that in the vicinity of ferromagnetism, 
a novel type of superconductivity with parallel-spin pairs 
(a spin-triplet state) is induced by FM fluctuations\cite{fujimoto-JPSJ73}.  
This superconductivity is robust in rather high magnetic fields, 
as the Pauli depairing \cite{clogston-PRL9} is irrelevant to Cooper pairs of parallel spins.  
Even in this state, however, $H_{c2}$ is still limited by orbital depairing \cite{WHH}. 
A novel mechanism is required to overcome the orbital limit and 
stabilize superconductivity in extremely high magnetic fields. 

The FM superconductor UCoGe, discovered by Huy {\it et al.} in 2007\cite{huy-PRL99}, allows us to study 
experimentally the electronic state near a FM critical point 
because of its low Curie temperature $T_{\rm Curie} \simeq 3$ K 
and small ordered moment $m_{0} \simeq 0.07 \mu_{B}$. 
Superconductivity sets in at $T_{\rm SC} \simeq 0.8$ K, which is the highest among FM superconductors discovered so far. 
Below $T_{\rm SC}$, ferromagnetism microscopically coexists with superconductivity, 
which was shown by $^{59}$Co nuclear quadrupole resonance experiments\cite{ohta-JPSJ77,ohta-JPSJ79}. 
That its $H_{c2}$ was greater than the Pauli-limiting fields by nearly one order of magnitude 
suggested a possibility of spin-triplet superconductivity\cite{huy-PRL100,slooten-PRL103}.  
Moreover, when magnetic fields are applied exactly parallel to the $b$ axis,
superconductivity is enhanced \cite{aoki-JPSJ78} 
and the $H_{c2}(T)$ curve shows an upturn at low $T$ to exceed the orbital limiting field.  
Similar strong superconductivity against fields was observed in the sister compound URhGe \cite{levy-NP3}.  
The enhancement of superconductivity was attributed to an increased effective mass 
in the vicinity of a ferromagnetic critical point \cite{levy-NP3,aoki-JPSJ78}, 
which is achieved by suppression of ferromagnetism by fields.  

In general, however, magnetic fields stabilize ferromagnetism and suppress ferromagnetic fluctuations. 
As the $H_{c2}$ enhancement occurs only in the filed along certain direction, 
anisotropic magnetism could be responsible for the unusual properties in fields. 
In fact, strong Ising anisotropy is observed in magnetization of both UCoGe \cite{huy-PRL100} and URhGe \cite{aoki-Nature413}. 
In addition to the static susceptibility probed by magnetization measurement, 
study on the anisotropy of dynamic susceptibility is crucial for sorting out the SC mechanism, 
because dynamic part is directly related to the magnetic fluctuations which induce superconductivity.

\begin{figure}[tbp]
\begin{center}
\includegraphics[width=7cm]{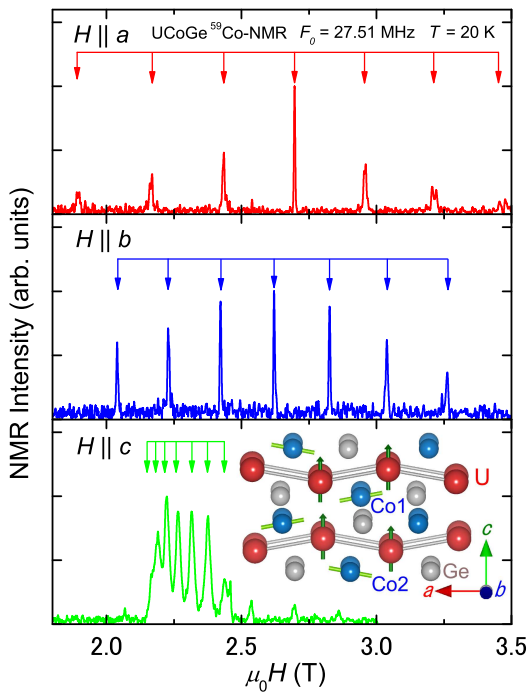}
\end{center}
\caption{
$^{59}$Co NMR spectra for fields along the three crystalline axes. 
$^{59}$Co nucleus has $I$ = 7/2, and thus seven peaks are observed for one Co site in the presence of EFG, 
which are shown with arrows. 
Inset : Crystal structure of UCoGe. 
The single crystallographic Co site becomes two inequivalent sites (Co1, Co2) under magnetic field.
Arrows at each Co site represent the principal axis of the EFG, 
and those at U site represent the magnetic easy axis. 
}
\label{fig1}
\end{figure}

We have performed $^{59}$Co NMR experiment on a single crystalline sample of UCoGe 
to investigate the direction-dependent magnetic properties in the normal state. 
The nuclear quadrupole splitting of the $^{59}$Co NMR spectra allowed us to determine the field directions precisely. 
Direction dependences of the Knight shift ($K$) and nuclear spin-lattice relaxation rate ($1/T_1$) 
provide clear microscopic evidence for Ising-type FM fluctuations both in the static and dynamic susceptibilities.  

We used the same $1.65 \times 1.65 \times 1.89$ mm$^3$ single crystalline sample 
grown using the Czochralski method \cite{huy-PRL100} 
as in the previous Co-NQR measurements \cite{ohta-JPSJ79}. 
$T_{\rm Curie}$ and the midpoint of $T_{\rm SC}$ in zero field are $2.5$ K and $0.57$ K, 
respectively \cite{ohta-JPSJ79, deguchi-JPSJ79}. 
Clear specific-heat jumps at $T_{\rm Curie}$ and $T_{\rm SC}$ attest to the high quality of the sample. 
The orientations of the external magnetic fields were 
carefully controlled {\it in situ} using a single-axis rotator. 
The other axis was aligned by eye and the misalignment is estimated as 
less than 3 degrees from detailed NMR spectrum analyses shown later. 
Three representative $^{59}$Co NMR spectra with fields along $a$, $b$ and $c$ directions are 
displayed in Fig.~\ref{fig1}. 

When a nucleus with a spin larger than unity sits at a position where the electric field gradient (EFG) is finite, 
the nuclear quadrupole interaction splits the NMR spectrum. 
The external field's angle with respect to the principal axis of the EFG is identified 
from the quadrupole-split NMR spectrum of the nucleus.   
$^{59}$Co nuclei ($I = 7/2$) in the single crystalline sample are suitable for angle-resolved NMR measurements.  
Since the EFG parameters for the Co site have already been determined from the previous $^{59}$Co NQR experiments\cite{ohta-JPSJ77}, 
the angle between the applied field ($H$) and the EFG axis can be determined from the $^{59}$Co NMR spectra.  
However, the angle between the EFG principal and crystalline axes  
cannot always be determined straightforwardly. 
We determined the field direction with respect to the crystalline axes of UCoGe  
by taking advantage of the local symmetry at the Co site. 
In UCoGe, the crystallographically unique Co site becomes two inequivalent sites under magnetic fields 
(Co1 and Co2 in the inset of Fig.~\ref{fig1}). 
In this case, the $^{59}$Co NMR spectrum consists of 14 peaks, seven for each site. 
In fields along the high symmetry axes, 
the pairs recombine and only seven peaks can be observed, 
as the two Co sites become equivalent again.  
The clear seven-peak spectra shown in Fig.~\ref{fig1} confirm the excellent field alignment. 
The relative angle between the EFG axes and crystalline axes can then be determined, 
as the field's angle with respect to the EFG is obtained for the seven-peak NMR spectra 
of $H \parallel a$, $b$, and $c$.  
These analyses allowed us to conclude that 
the $x$ axis of the EFG is identical to the crystalline $b$ axis, and 
the $z$ axis of the EFG is rotated by $10\pm2$ degrees from the crystalline $a$ axis. 
Note that the EFG principal axes were defined such that the EFG components order as $V_{zz} > V_{xx} > V_{yy}$. 
This result is in good agreement with theoretical results derived from 
a first-principle band calculation\cite{harima-unpublished}. 

\begin{figure}[tbp]
\begin{center}
\includegraphics[width=8cm]{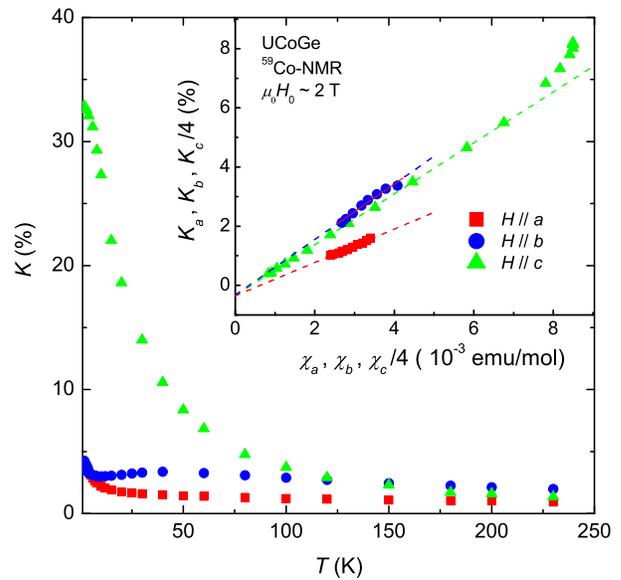}
\end{center}
\caption{
NMR shift $K$ measured in fields along the $a$, $b$, and $c$ directions. 
Remarkable anisotropy was observed. 
In the inset, $K$ is plotted against the bulk susceptibility $\chi$ measured in $2$ T. 
The nearly identical slops for all the directions indicate 
that the hyperfine coupling constants are isotropic.   
}
\label{fig2}
\end{figure}

Magnetic properties were measured in fields along the $a$, $b$, and $c$ axes. 
A huge Knight shift was observed only for $H \parallel c$, 
in agreement with bulk susceptibility results \cite{huy-PRL100}.  
The magnetic easy axis is represented by arrows at the U site in Fig~\ref{fig1}. 
In general, the NMR shift $K_{\alpha}$ for the field along the $\alpha$ direction is written 
in terms of the hyperfine coupling constant $A_{\rm hf}^{\alpha}$ and static susceptibility $\chi(\bm{q}=0)$ as
\begin{equation}\label{shift}
K_{\alpha} = A_{\rm hf}^{\alpha}\chi ^{\alpha}(0) + K^{\alpha}_{\rm orb}, 
\end{equation} 
where $K_{\rm orb}$ is temperature independent orbital contribution. 
The hyperfine coupling constant is determined by plotting $K$ against $\chi$ with $T$ as an implicit parameter. 
This $K$ versus $\chi$ plot, displayed in the inset of Fig.~\ref{fig2}, 
indicates that $A_{\rm hf}^{\alpha}$ is positive and roughly independent of the crystalline axis. 
The coupling constant can become isotropic when U-$5f$ electrons are transfered to the Co-$4s$ orbitals 
and interact with the Co nuclei directly. 
The anisotropic couplings, such as a dipole term, are found to have minor contributions. 
We also point out that the Co-$3d$ orbital, which gives rise to a negative hyperfine coupling constant, 
does not appear at the Fermi energy. 
The nearly isotropic $A_{\rm hf}$ and strongly anisotropic $K$ indicate that 
the U-$5f$ spins themselves give rise to the anisotropy due to the spin-orbit interaction. 
The spin part of the NMR shift $K^{\alpha}_{\rm spin}$ in the paramagnetic state is obtained 
by subtracting the orbital contribution from the total shift.  
The enormous anisotropy factor for static spin susceptibility $K_{\rm spin}^{c}/K_{\rm spin}^{ab}$, 
which reaches up to $10$ at low temperatures, 
clearly demonstrates that the static susceptibility has Ising anisotropy. 

\begin{figure}[tbp]
\begin{center}
\includegraphics[width=7cm]{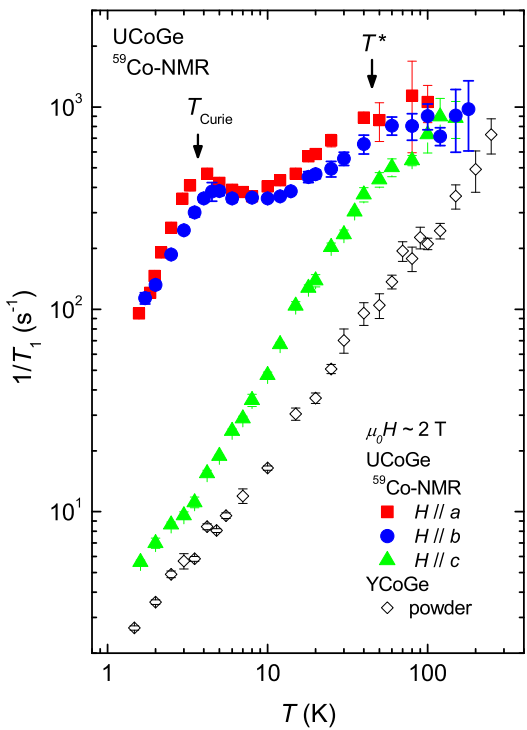}
\end{center}
\caption{ 
Nuclear spin-lattice relaxation rate ($1/T_{1}$) measured in three different field directions. 
The results of YCoGe, a reference compound without $f$ electrons, 
are also displayed. 
A broad peak observed at $4$ K shows the ferromagnetic transition broadened by fields. 
At $T$ higher than $T^*$, the Ising anisotropy weakens and $1/T_{1}$ becomes almost angle independent. 
}
\label{fig3}
\end{figure}

The nuclear spin-lattice relaxation rate $1/T_{1}$, measured for fields along the three crystalline directions down to $1.5$ K, 
is shown in Fig.~\ref{fig3}  
along with the result on YCoGe, a reference compound without $f$ electrons, is also displayed.   
The $1/T_{1}$ of YCoGe is proportional to $T$, 
which is observed in conventional metals and is known as Korringa behavior.  
The $5f$ electrons in the U compound induce magnetic scattering and add magnetic contributions to $1/T_{1}$. 
At temperatures higher than $80$ K, $1/T_{1}$ in UCoGe saturates to show $T$ independent behavior. 
In this region, the nuclear relaxation is dominated by the fluctuations of $5f$ localized moments, 
which exhibit Curie-Weiss behavior. 
The contribution from conduction electrons is estimated from $1/T_1$ for YCoGe, 
and found to be less than $1/5$ of the magnetic contribution. 
This $1/T_1 =$ constant behavior is commonly observed in heavy Fermion compounds far above coherence temperature ($T^*$), 
where the hybridization between localized moments and conduction electrons is weak. 
At temperatures below $T^*$, where local moments strongly hybridize with conduction electrons and 
have itinerancy with heavy mass, 
nuclear relaxation becomes anisotropic reflecting the anisotropic character of U-$5f$ electrons. 
This result indicates that the Fermi surface modification by the conduction and $f$ electrons hybridization 
is substantial. 

\begin{figure}[tbp]
\begin{center}
\includegraphics[width=8cm]{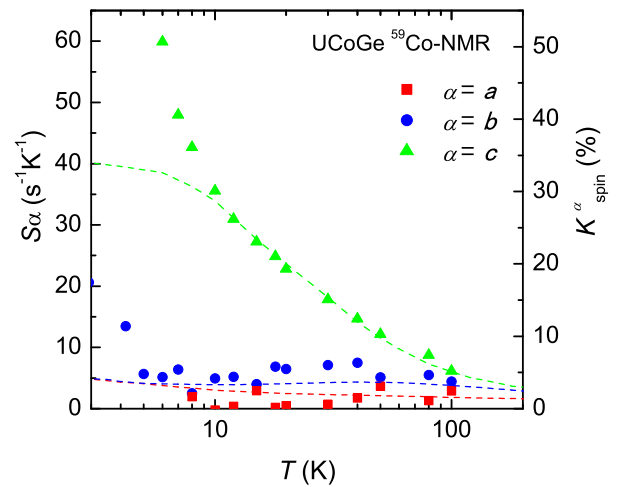}
\end{center}
\caption{
Direction-decomposed dynamic susceptiblity $S$ (see text) and 
static susceptibility $K_{\rm spin}$ along each direction. 
Identical Ising anisotropy for both quantities found above $8$ K 
suggests that the longitudinal mode dominates these fluctuations.  
It is noteworthy that in this $T$ range, the Knight shift scales with $S$, 
which is predicted by SCR theory for a nearly ferromagnetic metal. 
}
\label{fig4}
\end{figure}

In the low $T$ state, $1/T_{1}$ for $H \parallel c$ is nearly one order of magnitude smaller 
than those measured in the other two field directions. 
This angle dependence is in good contrast to that of the Knight shift, 
which is largest for $H \parallel c$.  
This is because the Knight shift probes the static susceptibility along the external field direction,  
whereas $1/T_{1}$ detects the fluctuations of the hyperfine fields $\delta H$ perpendicular to the fields. 
$1/T_1$ measured in a field along the $\alpha$ direction is expressed 
in terms of the imaginary part of the dynamic susceptibility along the $\beta$ and $\gamma$ directions, 
perpendicular to $\alpha$,  
$\chi''_{\beta, \gamma}({\bm q},\omega_0)$ as  
\begin{equation}
\label{relaxation} 
\left(\frac{1}{T_1T}\right)_{\alpha}  = 
\frac{\gamma_n^2 k_B}{(\gamma_e\hbar)^2} \sum_{{\bm q}} \left[
|A^{\beta}_{\rm hf}|^2\frac{\chi''_{\beta}({\bm q},\omega_0)}{\omega_0} + 
|A^{\gamma}_{\rm hf}|^2\frac{\chi''_{\gamma}({\bm q},\omega_0)}{\omega_0}\right], 
\end{equation} 
where $\gamma_n$ and $\omega_0$ are the gyromagnetic ratio and NMR frequency, respectively.  
With this equation and $1/T_{1}T$  measurement in three different field orientations, 
each term represented by $\sum_{{\bm q}}\left[|A_{\rm hf}^{\alpha}|^2\chi''_{\alpha}({\bm q},\omega_0)/\omega_0 \right] 
\equiv S_{\alpha}$ is separated out experimentally, 
and the results are shown in Fig.~\ref{fig4},
where the uniform spin susceptibility $K_{\rm spin}^{\alpha}$ along each axis is also shown by the dashed lines. 
In this plot, we found the Ising anisotropy with the easy axis along the $c$ axis also in the dynamic susceptibility $S_{c}$. 
The identical direction dependence in both static and dynamic susceptibilities reveal that 
the dominant magnetic fluctuations are along the magnetic easy axis (longitudinal mode). 
In addition, it is noteworthy that $S_{c}$ scales linearly with $K_{\rm spin}^{c}$ above $8$ K, 
indicative of the predominance of the FM fluctuations, 
as this scaling is anticipated for three-dimensional (3-D) FM fluctuations 
on the basis of self consistent renormalization (SCR) theory\cite{moriya-JMMM100}. 
 
Below $8$ K down to $T_{\rm Curie}$, $S_{c}$ shows an abrupt increase, 
which deviates from the temperature dependence of $K_{\rm spin}^{c}$. 
This is due to the difference in field direction with respect to the probed susceptibility; 
$K_{\rm spin}^c$ was measured in $H \parallel c$, while  $S_{c}$ was obtained in $H \perp c$.
The critical fluctuations in the vicinity of this FM transition are easily suppressed by fields pointing along the $c$ axis, 
while in the other field directions these fluctuations can survive down to $T_{\rm Curie}$ as seen in Fig.~\ref{fig3}.  
It should be noted that $1/T_1$ in $H \parallel c$ does not show any anomaly around $T_{\rm Curie}$. 
This suggests that the dynamic susceptibilities along the $a$ and $b$ axes are not affected 
by the critical FM fluctuations at all, which is also consistent with Ising-type FM fluctuations.

Now, we discuss the possible SC pairing state induced by the magnetic fluctuations revealed above. 
It has been reported \cite{fay-PRB22} that the FM fluctuations in the vicinity of ordering mediate spin-triplet superconductivity. 
This is consistent with the large $H_{c2}$ exceeding the Pauli limiting field \cite{huy-PRL99}.
In addition, Monthoux and Lonzarich\cite{monthoux-PRB59}, and Fujimoto\cite{fujimoto-JPSJ73} have pointed out that 
$T_{\rm SC}$ for Ising FM fluctuations is greater than that for isotropic FM fluctuations under similar conditions. 
In the case of the Ising FM fluctuations with only the longitudinal mode, 
pair breaking caused by the transverse spin fluctuations can be minimized 
even in the vicinity of the quantum critical point. 
The longitudinal fluctuations found in UCoGe could consistently 
explain the pressure-temperature phase diagram \cite{slooten-PRL103}, 
where $T_{\rm SC}$ shows a maximum at the FM quantum critical point, 
and SC phase extends to the outside of the FM phase.  

In conclusion, 
we have performed angle-resolved $^{59}$Co NMR measurements on single crystalline UCoGe. 
The static and dynamic susceptibilities along each crystal axis derived from Knight-shift and $1/T_1$ measurements 
show the presence of Ising-type FM fluctuations along the magnetic easy axis ($c$ axis) above $1.5$~K.   
The FM critical fluctuations around $T_{\rm Curie}$ are not significantly suppressed when magnetic fields are applied perpendicular to the $c$ axis. 
If superconductivity is induced by these Ising FM fluctuations, 
spin-triplet state is proposed theoretically. 
Then, the large $H_{c2}$ and extended SC phase in pressure-temperature phase diagram can be consistently explained.  
Experimental evidence to reveal the relationship between magnetic fluctuations and superconductivity 
is required, as the next step, to understand the FM SC state. 

The authors thank D. C. Peets, S. Yonezawa, and Y. Maeno for experimental support and valuable discussions, 
and H. Harima, H. Ikeda, S. Fujimoto, Y. Tada, A. de Visser, D. Aoki, and J. Flouquet for fruitful discussions. 
This work was partially supported by Kyoto Univ. LTM center, 
the ``Heavy Electrons'' Grant-in-Aid for Scientific Research on Innovative Areas  (No. 20102006) from MEXT of Japan, 
a Grant-in-Aid for the Global COE Program 
``The Next Generation of Physics, Spun from Universality and Emergence'' from MEXT of Japan, and
a Grant-in-aid for Scientific Research (No. 20340088) from JSPS and KAKENHI (S) (No. 20224015) from JSPS.

\end{document}